\documentclass{emulateapj}

\shorttitle{The Monoceros ring in the Anticenter}
\shortauthors{Sollima et al.}

\begin{document}


\title{A deep view of the Monoceros ring in the Anticenter direction: 
clues of its extra-Galactic origin}


\author{A. Sollima\altaffilmark{1,2,3}, D. Valls-Gabaud\altaffilmark{4},
D. Martinez-Delgado\altaffilmark{5,1,2}, 
J. Fliri\altaffilmark{4},  J. Pe{\~n}arrubia\altaffilmark{6}, H.
Hoekstra\altaffilmark{7,8}}
%
%
\email{antonio.sollima@oapd.inaf.it}


\altaffiltext{1}{Instituto de Astrofisica de Canarias, E38205 San Cristobal de La Laguna, E38205, Spain}
\altaffiltext{2}{Departamento de Astrofísica, Universidad de La Laguna, E-38205
San Cristobal de La Laguna, Spain}
\altaffiltext{3}{INAF Osservatorio Astronomico di Padova, I35122 Padova, Italy}
\altaffiltext{4}{GEPI, CNRS UMR 8111, Observatoire de Paris, 92195 Meudon, France}
\altaffiltext{5}{Max Plank Institut f\"{u}r Astronomie, D69117 Heidelberg, Germany}
\altaffiltext{6}{Institute of Astronomy, University of Cambridge, Cambridge, CB3 0HA, United Kingdom}
\altaffiltext{7}{Department of Physics and Astronomy, University of Victoria, Victoria, BC V8P 1A1, Canada}
\altaffiltext{8}{Leiden Observatory, Leiden University, PO Box 9513, 2300 RA,
Leiden, the Netherlands}


\begin{abstract}
We present the results of deep imaging obtained at the CFHT with MegaCam  
in the Anticenter direction at two different heights above the Galactic disk. 
We detect the presence of the Monoceros ring in both fields as a conspicuous 
and narrow Main Sequence feature which dominates star counts over a 
large portion of the color-magnitude diagram down to $g'\sim24$.
The comparison of the morphology and density of this feature with a large
variety of Galactic models excludes the possibility that it can be due to a
flare of the Galactic disk, supporting an extra-Galactic origin for this
ring-like structure.
\end{abstract}





\section{Introduction}

The Monoceros ring represents one of the most controversial features among
the various sub-structures discovered in the Milky Way.
It was discovered as a ring-like structure spanning about $170^\circ$ in 
longitude at a nearly constant Galactocentric distance at low-latitudes 
(Newberg et al. 2002; Yanny et al. 2003).
Its structure and extent have been later studied by many authors using both
photometric (Rocha-Pinto et al. 2003; Ibata et al. 2003; Conn et al. 2005a, 2007)
and spectroscopic (Crane et al. 2003) data. On the basis of these analyses, the Monoceros (Mon)
stream has been found on both sides of the plane of the Galaxy at 
Galactocentric distances $15<R<20$ kpc. 
Subsequent spectroscopic studies indicated that this structure is kinematically
cold ($13<\sigma_{v}<20$ Km s$^{-1}$; Conn et al. 2005b; Martin et al. 2006;
Casetti-Dinescu et al. 2008) and homogeneous in metal content
($\sigma_{[Fe/H]}\sim 0.15~dex$; Ivezic et al. 2008, hereafter I08). 
The nature of the Mon ring is highly debated: in fact it is not clear if 
this feature is a debris from an accreted satellite (e.g. Helmi et al.
2003; Martin et al. 2004; Penarrubia et al. 2005; Martinez-Delgado et al.
2005) or an intrinsic feature of the Galactic disk possibly associated to the
disk flare (Momany et al. 2006, hereafter M06; Hammersley \& Lopez-Corredoira
2011), the Norma-Cygnus spiral arm (Moitinho et al.
2006), perturbations due to past accretion events (Kazantzidis et al. 2008) or 
the second dark matter caustic ring (Natarajan \& Sikivie 2007).

In this {\sl Letter} we present the detection of the Mon ring toward the
Galactic anticenter at two different Galactic latitudes and use its
color-magnitude diagrams (CMDs) to investigate the possible connection between this 
object and the Galactic disk flare.
The Anticenter represents a particularly convenient direction to test this
scenario since {\it i)} it lies close to the line of nodes where the effects of the
Galactic warp are minimized (Drimmel \& Spergel 2001), {\it ii)} it avoids
contamination from bulge stars, and {\it iii)} it lies
along a radial away from the Galactic center, following the
direction of growth of the disk scale-height. 
This analysis comes from two photometric campaigns performed
with MegaCam@CFHT devoted to the study of weak lensing in massive clusters of
galaxies (see Hoekstra 2007)
and to the search for extra-tidal structures in the 
outskirts of outer halo globular clusters (Martinez-Delgado et al. 2004).

\section{Observations and Data reduction}

The photometric data have been obtained with the MegaCam camera, mounted at the
Canada-France-Hawaai Telescope (CFHT)(Mauna Kea, Hawaai). 
The camera consists of a mosaic of 36 chips with a pixel scale of 0.185" 
$pixel^{-1}$ providing a
global field of view of $\sim~1^{\circ}\times 1^{\circ}$. Observations have been 
performed in Service Mode in two different observing runs on October 2005 (3
nights) and January 2009 (two nights).
The two datasets consists in a set of 15 (4 $g'$ and 11 $r'$) 600 sec-long and
13 (6 $g'$ and 7 $r'$) 680 sec-long images centered at 
$(\alpha,\delta)=(7h~18m~08s,+37^{\circ}~37'~45")$(field 1) and
$(\alpha,\delta)=(7h~38m~08s,+38^{\circ}~54'~00")$(field 2)
with a dithering pattern of few arcminutes to fill the gaps
between the chips. 
The average seeing for the two datasets was 0.7" and 1.0", respectively.

The standard reduction steps (bias, dark and flat-field correction) have been
performed using the Elixir pipeline developed by the CFHT team.
We used DAOPHOT II and the PSF fitting algorithm 
ALLSTAR (Stetson, 1987) to obtain instrumental magnitudes for all the stars 
detected in each frame. 
Mean frames have been obtained by aligning and averaging all the $g'$ and $r'$ 
images with a 3 $\sigma$ clipping rejection thresold, using the 
Terapix\footnote{http://terapix.iap.fr} software specifically devoted to this
task. 
The automatic detection of sources has been performed on the mean frame adopting a 
3$\sigma$ treshold. The file with the object positions has been then used as input for the
PSF-fitting procedure, that has been performed independently on each image. As
usual, the most isolated and bright stars in each field have been employed to
build the PSF model (here a Moffat function of exponent 2.5 has been adopted).
For each passband, the derived magnitudes have been transformed into
the same instrumental scale and averaged. We adopted the nightly zero points 
and reddening coefficients provided by the CFHT team to link the
instrumental magnitudes to the standard system.
Finally, a catalog with more than 40,000 calibrated sources has
been produced. 
The photometrically calibrated catalog has been cross-correlated with the
seventh data release of the Sloan Digital Sky Survey (SDSS; Abazajian et al. 2009) catalog, which lists accurate
positions for more than 3,000 objects over an area of 
$\pi$ sq.deg. around both fields, to derive an accurate astrometric
solution with a typical r.m.s of 200 mas.
These catalogs represent the deepest and most accurate 
datasets on the Mon feature, with over 3 times more stars and a photometric
accuracy about an order of magnitude better than previous SDSS data.

\section{Results}

\subsection{Color-Magnitude Diagrams}

\begin{figure}
\epsscale{1.2}
\plotone{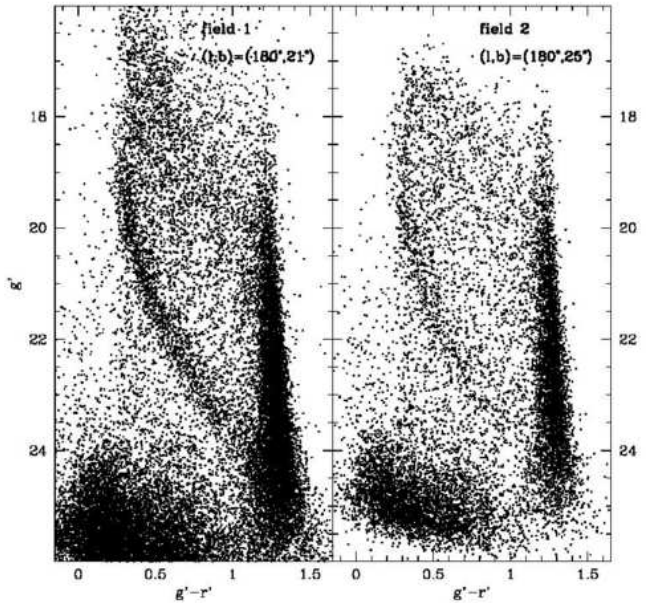}
\caption{CMDs of field 1 (left panel) and field 2 (right panel). 
Only stars with a sharpness parameter 
$\vert S \vert < 0.5$ are plotted. The clump visible at $g'>24$, $g'-r'<$0.5 is
due to the contamination of unresolved background galaxies.}
\label{cmd}
\end{figure}

In Fig. \ref{cmd} the $g'-r',~g'$ CMDs of the two fields are shown. 
Only objects with a sharpness parameter $\vert S
\vert<$0.5 (as defined by Stetson 1987) have been plotted to avoid 
contamination from background galaxies. 
As can be noted, the CMDs sample the foreground Galactic population, 
reaching a limiting magnitude $g'\sim 25.5$. 
A significant overdensity of stars can be noticed in both fields at
$g'-r'<1$ and $19<g'<24$, with a
morphology which resembles a Main Sequence (MS). 
It is evident from Fig. \ref{cmd} that the density of MS stars is significantly
smaller in field 2 compared with field 1.

The present detections are located close to the region where the Mon ring was
discovered (Newberg et al. 2002). More recently, Grillmair (2006) found a compact overdensity of stars above the
background field population at an height of $\sim$30 degrees above the Galactic plane 
close to the previous detection of Mon (the "Anticenter stream"). Although the possible association
between the Anticenter stream and Mon is still debated (see Carlin et al. 2010 and references therein), our target fields
are located at lower Galactic latitudes ($\Delta b\sim 12^{\circ}$) with respect to the 
Anticenter stream overdensity, being therefore unrelated with such structure.

\subsection{Synthetic CMDs}

\begin{table}
\begin{center}
\caption{Best-fit parameters of the best-fit S11 model \label{tbl-1}}
\begin{tabular}{lcl}
\tableline\tableline
Parameter & Value & Description\\
\tableline
$R_{\odot}$	     & 8.5 kpc  & Solar Galactocentric distance\\
$Z_{\odot}$	     & 15 pc    & Solar height above the Galactic plane\\
$q$                  & 0.63     & Halo flatness parameter\\
$f$		     & 0.9      & Fraction of thin disk stars\\
$f_{h}$ 	     & 0.001    & Fraction of halo stars\\
$n_{h}$ 	     & 2.77     & Halo density power-law coefficient\\
$R_1$		     & 1.99 kpc & Thin disk scale-length\\
$R_2$		     & 2.99 kpc & Thick disk scale-length \\
$h_{Z,1}(R_{\odot})$ & 193 pc   & Thin disk scale-height at the Solar circle\\
$h_{Z,2}(R_{\odot})$ & 611 pc   & Thick disk scale-height at the Solar circle\\
$h_{fl}$	     & 23.0 kpc & Flare characteristic radius\\
$Z_{0}$ 	     & 370 pc   & Maximum warp height\\
$R_{w}$ 	     & 11.5 kpc & Warp starting radius\\
$h_{w}$ 	     & 3.0 kpc  & Warp characteristic radius\\
$\phi_{0}$	     & $11^{\circ}$ & Line of nodes angle\\
\tableline
\end{tabular}
\end{center}
\end{table}
\label{s_analys}

\begin{figure}
\epsscale{1.2}
\plotone{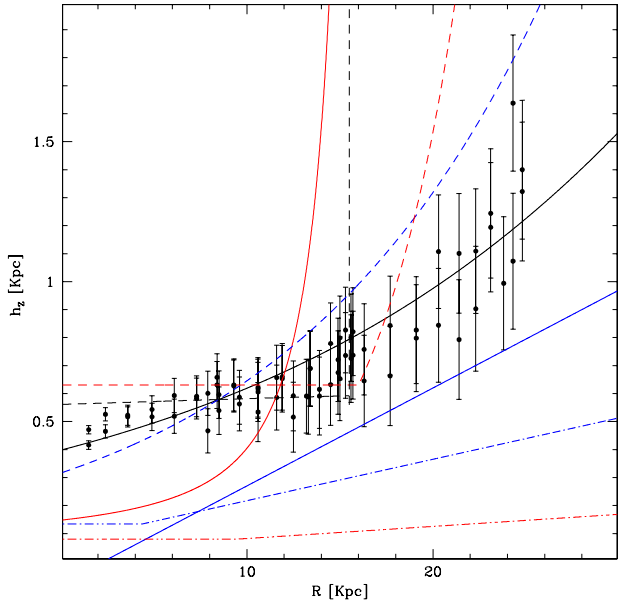}
\caption{Disk scale-height as a function of the Galactocentric
distance. The measures by M06 are plotted as filled circles. The prediction of 
the various models (DS01: blue dot-dashed line; L02: red solid line; R03: 
red dot-dashed line; FR03: blue solid line; Y04: blue dashed line; S11: black
solid line; HL11: red
dashed line; "wall flare" model: black dashed line) are overplotted.}
\label{model}
\end{figure}

\begin{figure*}
\epsscale{1.}
\plotone{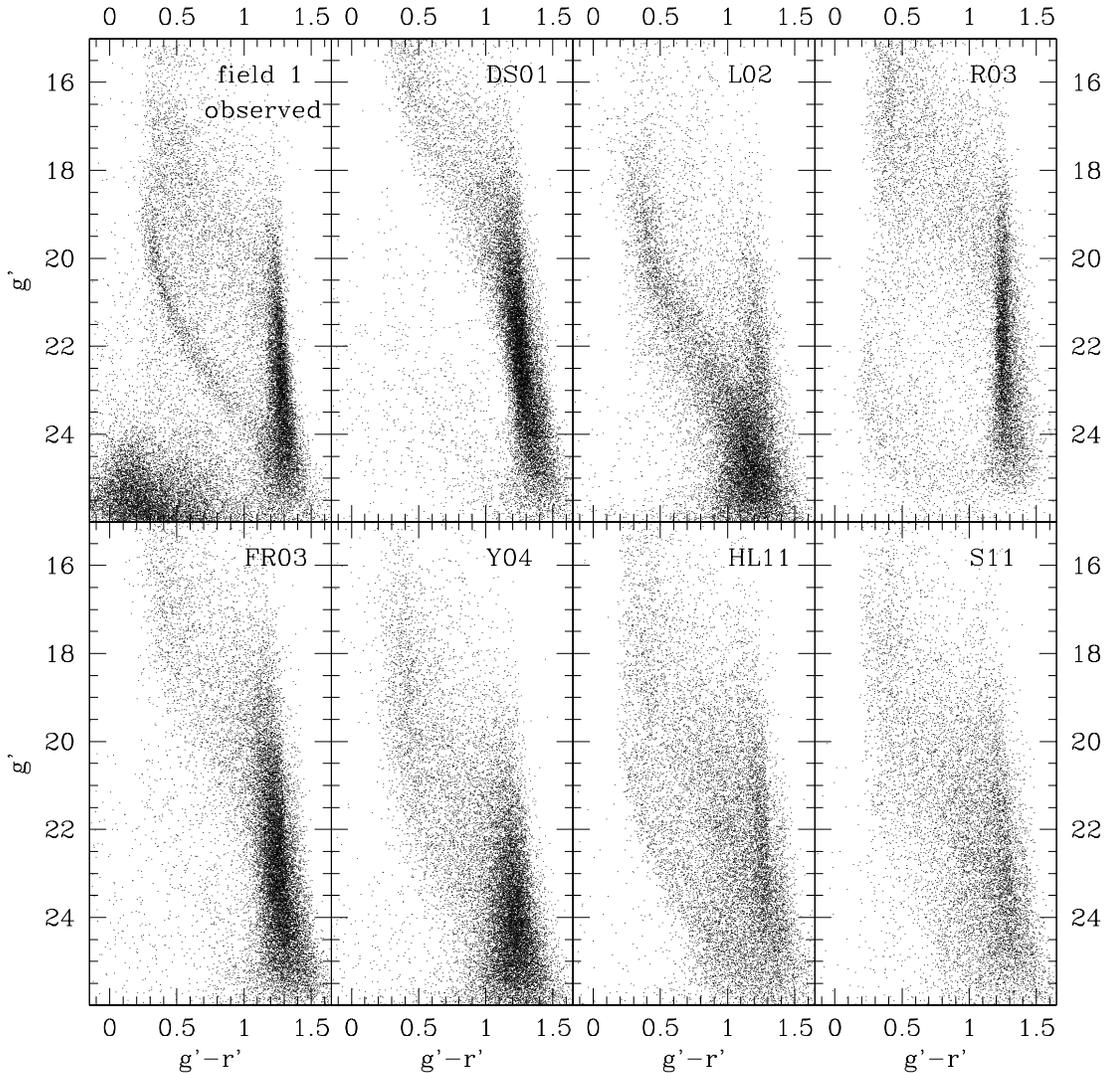}
\caption{Comparison between the CMDs of field 1 and the seven Galactic models
considered in this work.}
\label{field}
\end{figure*}

\begin{figure*}
\epsscale{1.}
\plotone{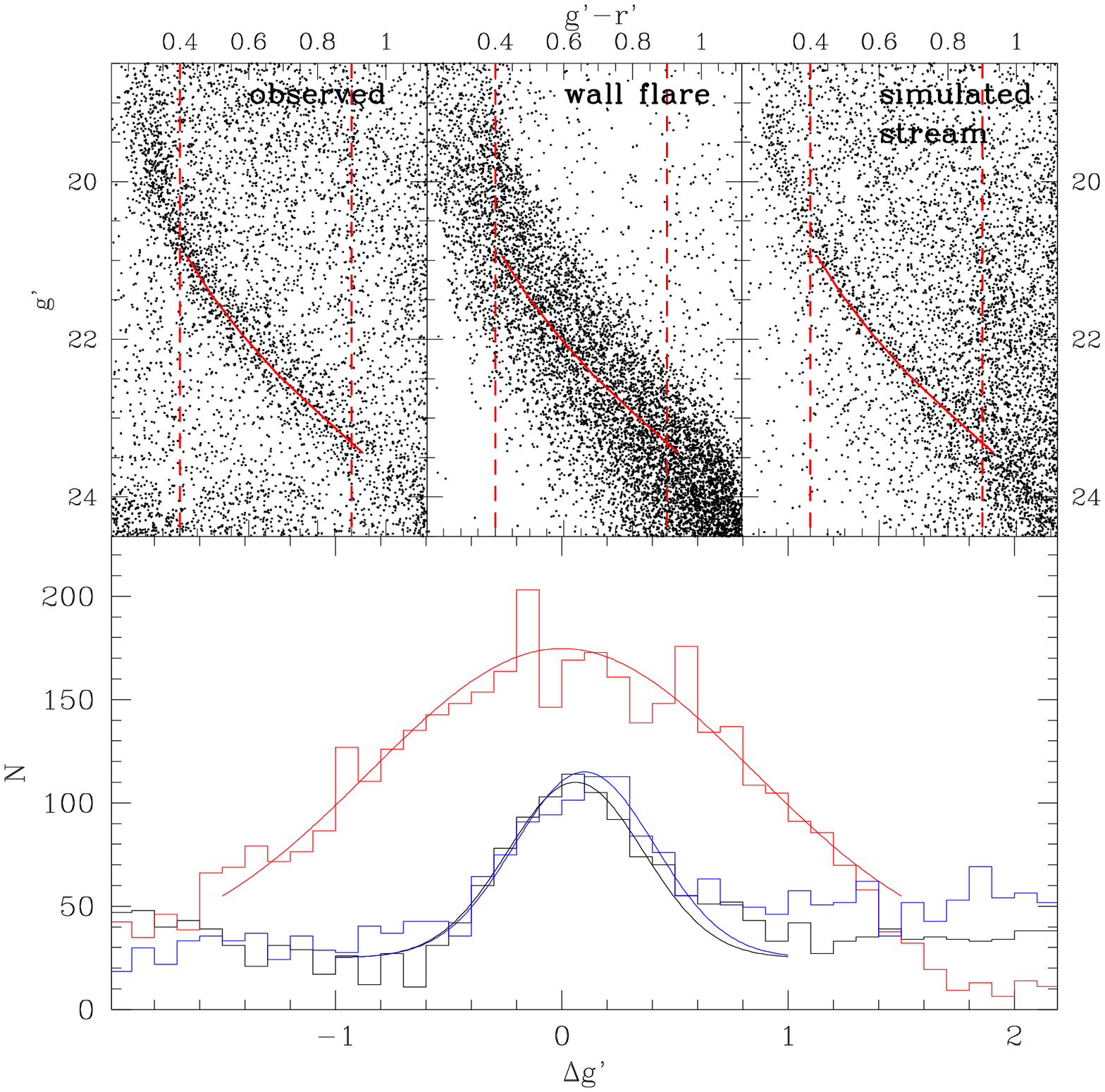}
\caption{Upper panels: Zoomed CMDs of field 1 (upper-left panel), "wall flare" model
(upper-central panel) and simulated stream model (upper-right panel). The MS
ridge line and the color range used to construct the magnitude distribution are
shown with red solid and dashed lines, respectively. Bottom panel: Distribution of magnitude differences about the MS ridge line for the
observed CMD of field 1 (black histograms), "wall flare" model (red histograms) and the
simulation for a discrete stream (blue histograms). The best-fit gaussians are
overplotted.}
\label{msfit}
\end{figure*}

To test the possible association of the MS feature observed in Fig. \ref{cmd}
with the disk flare we simulated a series of synthetic CMDs for the 
Galactic field population following the prescription of seven different Galactic 
models: Drimmel \& Spergel (2001; hereafter DS01); Lopez-Corredoira et al.
(2002; L02); Robin et al. (2003; R03), Foster \& Routledge (2003; FR03),
Yusifov (2004; Y04), Hammersley \& Lopez-Corredoira (2010; HL11) and 
a simple model which has been calibrated
to fit the 2MASS data of the disk flare and warp collected by M06. 
Hereafter, we will refer to this best-fit model as S11. 
For this last model we adopted an oblate spheroidal halo with a declining
profile of the form
\begin{equation}
\rho_{h}(R,Z)=f_{h}~\rho_{\odot,d}~\left(R_{\odot}/\sqrt{R^{2}+(Z/q)^2}\right)^{n_{h}}
\end{equation}
The disk has been modelled as a sum of two exponential disks (Juric et al. 2008; hereafter J08)
\begin{equation}
\rho_{d}(R,Z)=\rho_{\odot,d}~(f~e^{-(R-R_{\odot})/R_{1}-(Z+Z_{\odot})/h_{Z,1}}+
\end{equation}
$$(1-f)~e^{-(R-R_{\odot})/R_{2}-(Z+Z_{\odot})/h_{Z,2}})$$
The scale-heights of both disks have been assumed to increase with distance from the
Galactic center to take into account the disk flare
\begin{equation}
\label{eq_fl} 
h_{Z,i}(R)=h_{Z,i}(R_{\odot})~e^{(R-R_{\odot})/h_{fl}}
\end{equation}
Finally, the disk warp has been modelled as a series of tilted rings whose
disk heigths is defined as
\begin{equation} 
Z_{w}(Z,\phi)=Z_{0}~tanh((R-R_{w})/h_{w})~sin(\phi-\phi_{0})~~~\mbox{for } R>R_{w}
\end{equation}
where $\phi$ is the angle with respect to the Sun direction in a cylindrical system 
of coordinates centered on the Galactic center. 
In the definition of our best-fit model we kept fixed those parameters derived 
by J08 and left free the 4 ones which determine the warp and 
flare shape ($h_{fl},~Z_{0},~h_{w}$ and $\phi_{0}$). 
The final values of all the parameters involved in the above model 
are listed in Table 1.
The value of $\rho_{\odot,d}$ has been normalized to
reproduce the observed number of stars in field 1.

All the above models take into account for both the disk flare and
warp and have been calibrated to reproduce the density of different
markers: warm dust (DS01), bright stars (L02), dwarf stars (R03, HL11), 
HI gas (FR03), pulsars (Y04) and red clump stars (M06).
In Fig. \ref{model} the prediction of the various models
considered in this work are compared with the data collected by M06.
Note that significant differences are evident between the various models
mainly resulting from the different objects used as disk markers. 

The overall CMD has been simulated as the sum of the contribution
of the halo and disk stellar populations. We ignored the bulge contribution assuming it
negligible in the Anticenter direction.
For each Galactic component we randomly extracted a population of
stars from a Kroupa (2001) mass function with suitable ages and metallicities\footnote{Being the simulated CMDs mainly constituted by unevolved
stars, they are not affected by the uncertainties on the adopted SFR for the various Galactic components, but mainly on their metallicity distributions, which are well
constrained by spectroscopic observations (Venn et al. 2004).}. 
Only R03 and HL11 distinguish between thin and thick disk adopting two different
characteristic lengths and heights. For the other models
we simulated the Galactic disk population as a mixture of thin and thick disk 
stars in proportions estabilished by J08. This
approximation ensures a good level of accuracy in reproducing the CMD
(see also DS01; L02; FR03; M06; Reyle et al. 2009).
We adopted a two-step star formation rate (SFR) and the age-metallicity relation
from Fuhrmann (1998) for the thin disk (Girardi et al. 2005), while for the thick disk a 
constant SFR between 11 and 12 Gyr and a gaussian metallicity distribution with
$[M/H]=-0.4\pm 0.1$ has been assumed (Girardi et al. 2005).
For the halo, we assumed a constant SFR between 12 and 13 Gyr and the metallicity
distribution derived by Ryan \& Norris (1991).
A fraction of binaries of $f=50\%$ (Duquennoy \& Mayor 1991) has been simulated by randomly associating
pairs of stars and calculating the corresponding magnitudes by summing the
fluxes of the two components (see Rubenstein \& Bailyn 1997).
Absolute magnitudes and dereddened colors have been then derived using the 
evolutionary tracks by Marigo et al. (2008) in the MegaCam photometric system.
Halo and disk stars have been then located at different distances according to
the prescription of each Galaxy model.
For each star a proper extinction has been included by calculating the dust 
column density at the star's distance 
assuming a dust distribution across the disk of the form $\rho_{dust}\propto e^{-R/R_{dust}} sech^2(Z/Z_{dust})$ with a scale-heigth of
$Z_{dust}=134.4$ pc and a scale-length of $R_{dust}=2.26$ kpc (DS01), and normalized by imposing the extinction 
at infinity predicted by the Schlegel et al. (1998) maps.   
The distance and extinction have been finally used to convert absolute
magnitudes and colors into apparent ones.
Photometric errors and incompleteness have been estimated by means of artificial
stars experiments which have been performed on the CFHT images. A detailed
description of the adopted procedure can be found in Bellazzini et al. (2002). 

In Fig \ref{field} the observed CMD of field 1 and the corresponding synthetic 
CMDs are shown. It is evident that while models DS01, R03, FR03 and S11 well
reproduce the overall morphology of the Galactic field evolutionary sequences, none of them is able to reproduce
the prominent MS feature clearly visible in the observed CMD.
A poor representation of both the overall CMD and the Mon MS is instead 
given by Y04 and HL11 models.
Only model L02 predicts a compact distribution of stars resembling a MS. 
Such an increase is due to the extreme flaring of the disk adopted by L02 (see
Fig. \ref{model}) which produces a sudden increase of the density at a well 
defined Galactocentric distance (R$\sim$ 14 kpc), while the double-exponential density law 
($\rho_{d}\propto e^{-R/h_{R}-Z/h_{Z}}$) dampen the distribution of stars at
longer distances. 
Note however that this model (which has been extrapolated here outside its range of validity; $R<15~kpc$) 
fails to reproduce the magnitude spread of the observed MS feature as well as the morphology of both 
the red (at $g'<18.5,~g'-r'<0.7$) and blue (at $g'-r'>1$) field dwarfs in Fig. \ref{cmd}.

Summarizing, it seems impossible to reproduce the observed morphology of the CMDs with the existing models of disk flare. 
To check the general validity of such conclusion we compared the observed CMD with a model where the most extreme disk flaring has been assumed.
For this purpose we replaced in model S11 the functional form of the disks flaring (eq. \ref{eq_fl}) with a "wall" function where the scale-height 
of the two disks abruptly goes to infinity at a given
Galactocentric distance (set to 15.5 kpc to fit the mean magnitude of the Mon MS; see Fig. \ref{model}). The simulated CMD for this model (hereafter referred as "wall
flare") is shown in the central panel of Fig. \ref{msfit}. 
In Fig. \ref{msfit} we compared the distribution of magnitude differences about the MS ridge line in the color interval $0.4<g'-r'<0.9$. 
It is apparent that the "wall flare" model predicts a width of the MS which is not compatible with the observed one: the
standard deviation of the distribution is $\sigma_{\Delta g'}=0.30\pm0.03$ in the observed CMD
and $\sigma_{\Delta g'}=0.85\pm0.08$ in the "wall flare" model (i.e. $\sim$ 3 times larger).
Only an "ad hoc" abrupt cutoff of the density profile of both disks at a distance close to the flare radius (in contrast
with what found by J08) would reduce the magnitude spread of this model.
A second test has been performed by comparing the ratio between the number of 
MS stars observed in field 1 and 2 with the model prediction. For this purpose, 
we defined a region in the CMD
containing all stars with g' magnitude within 0.5 mag about the MS ridge line
defined in field 1, a color $0.4<g'-r'<0.9$ and a magnitude $g'<23.8$, to 
avoid contamination from Galatic dwarf stars and to ensure 
a level of completeness $\psi>90\%$ in both fields.
While the fraction of MS stars appears to increase by a factor
$N_{1}/N_{2}=1.87\pm0.11$, the "wall flare" model predicts a small decrease of this fraction
($N_{1}/N_{2}=0.92\pm0.03$)\footnote{The $N_{1}/N_{2}$ population ratio
calculated in the "wall flare" model turns out to be smaller than unity since
the small shift in magnitude produced by the decrease of latitude moves part of
the MS feature outside the adopted selction box. This is a consequence
of the large magnitude spread of the MS feature predicted by this model.}. 
Therefore, on the basis of the above analysis we conclude that even the most extremely acceptable flared disk model
cannot reproduce the observed CMD since it fails to predict: 
{\it i)} the population of both blue and red field dwarfs;
{\it ii)} the 2MASS data of the flare measured by M06 (see Fig.\ref{model}), 
{\it iii)} the magnitude dispersion of the observed MS feature, and 
{\it iv)} the ratio of MS stars at the two observed Galactic latitudes.

For comparison with the stream scenario, we constructed the synthetic CMD of a galaxy remnant located 
at a given distance to the Sun with a variable thickness. 
For this purpose, we apply the Bayesian formalism
developed by Hernandez \& Valls-Gabaud (2008) to infer the posterior probability 
distribution functions of distance and age, using as priors a gaussian 
function for metallicity, centred
on $[Fe/H]=-0.95$ and with dispersion 0.15 (I08).
We obtain maxima in the marginalised
posterior probability distribution functions at an age of $t=9.2\pm0.2$ Gyr
and a distance of $d=9.1\pm0.2$ kpc with a dispersion of 
$\sigma_{d}=0.90\pm0.08~kpc$, consistent with previous estimates for the
Mon stream (Newberg et al. 2002) and the predictions of Penarrubia et al. (2005). The obtained CMD is shown in 
the right panel of Fig. \ref{msfit}. In this last case the agreement with the observed CMD is
striking. 
A good agreement can be also found by assuming the Mon ring formed by 
stars with the same age and metallicity distribution of the Galactic disk (e.g. the case
of a disk perturbation or a spiral arm). In this case, the best-fit distance
and dispersion turn out to be $d=12.6\pm0.8~kpc$ and $\sigma_{d}=0.3\pm0.2~kpc$,
respectively. However, it is worth noting that while agreement can be had over 
the limited magnitude range used for estimating distance, the metallicity 
estimates by I08 argue against this scenario.

\section{Discussion}

In this {\sl Letter} we present the results of a deep MegaCam imaging of two regions
in the Galactic Anticenter direction where the presence of the Mon ring is
clearly evident. The high quality of our CMDs allowed to compare the morphology of the
observed MS feature with the predictions of the most recent Galactic models
including the effect of disk flare and warp. We find that none of the
considered models is able to reproduce the position and morphology of the
Mon MS or the ratio between the number of MS stars in the two
observed fields, even with extreme flares. We therefore conclude that, in the absence of an "ad hoc" abrupt disk cutoff, {\it the photometric signature of 
the Mon ring cannot be explained by any smooth variation of the Galactic
disk structure.} 
This result is in contrast with the conclusion drawn by HL11. The reason of
such a discrepancy stems from the method used by these authors to calculate the 
predicted star counts of their model: indeed, they adopted a delta function in 
magnitude to calculate the number of stars displaced at a given distance to the 
observer neglecting many important factors like the slope of MS stars 
in the CMD, the composite stellar population of the Galactic components 
and the photometric errors that all contribute to increase the 
magnitude spread of MS stars. By neglecting these effects HL11 interpreted the 
overall magnitude spread of the Mon MS as entirely due to a distance 
spread, overestimating its actual dispersion.

Of course, in absence of direct spectroscopic data, our data cannot exclude the 
other possible interpretations of the
Mon ring as the result of small scale perturbations of the Galactic disk due to various
possible physical processes (see e.g.   
Moitinho et al. 2006; Natarajan \& Sikivie 2007; Kazantzidis et al. 2008).
However, our CMDs impose strong constraints on the vertical structure and
extent of such a structure.
Indeed, if we assume the Mon ring formed by Galactic disk stars, the 
location in the CMD of the observed MS feature and its dispersion
in magnitude are compatible with an object located at a distance $d=12.6\pm0.8~kpc$ to the Sun
and a width of $FWHM < 1~kpc$. This corresponds to a very compact structure
located at an height above the Galactic plane of $Z\sim4.5~kpc$ (i.e. $\sim$ 4.5 
times the local scale-height of the disk). 
Moreover, the density of Mon stars decreases of almost a factor
of two by moving only $4^\circ$ toward higher Galactic latitudes (see Sect.
\ref{s_analys}). 
On the other hand, under the assumption that the Mon ring is due to an
accretion event which occurred in a past epoch, the derived distance of
$d=9.1\pm0.2~kpc$ and dispersion
$\sigma_{d}=0.90\pm0.08~kpc$ (FWHM=$2.1\pm0.2$ kpc) are in good
agreement with the prediction of the model by Penarrubia et al. (2005; see their
Fig. 7).

Our results lend support to the
previous studies on the metallicity distribution (I08), abundance patterns anomalies (Chou et al. 2010) and
kinematics (Casetti-Dinescu et al. 2008) of this object 
which show that the stellar composition of the
Mon ring is different from that of the Milky Way, favouring the extra-Galactic scenario. 
An accurate high-resolution spectroscopic survey of these faint MS stars would defitively
clarify the real nature of the Mon ring.

\acknowledgments

Based on observations obtained with MegaCam, a joint 
project of CFHT and CEA/DAPNIA, at the CFHT observing programs 05BC19 and 09AF03.
This work is based on data products produced at TERAPIX and the Canadian
Astronomy Data Centre.
This research was supported by the Spanish Ministry of Science and Innovation 
(AYA2007-65090), by ANR POMMME (ANR 09-BLAN-0228) and CNRS/MAE
PICASSO. We warmly thank the anonymous referee and M. Lopez-Corredoira for their helpful suggestions and Y. Momany for providing his data.

{\it Facilities:} \facility{CFHT}.


\end{document}